\documentclass[aps,prd,onecolumn,groupedaddress,nofootinbib,
superscriptaddress,preprint,11pt]{revtex4-1}
\pdfoutput=1
\usepackage[a4paper,margin=1in]{geometry}
\usepackage{color,xcolor}
\usepackage[colorlinks,citecolor=blue,urlcolor=blue,linkcolor=blue]{hyperref}
\usepackage{amsmath,amssymb}
\usepackage{graphicx}
\usepackage{slashed}
\usepackage{url}
\usepackage{enumitem}
\usepackage{multirow}
\usepackage{array}

\newcolumntype{?}{!{\vrule width 1pt}}
\begin{document}

\title{Improving Heavy Dijet Resonance Searches Using Jet Substructure at the LHC}

\preprint{HRI-RECAPP-2019-005}

%
%

\author{Aruna  Kumar  Nayak}
\email{nayak@iopb.res.in} 
\affiliation{Institute of Physics, HBNI, Sachivalaya Marg, Bhubaneswar 751\,005, India}

\author{Santosh Kumar Rai}
\email{skrai@hri.res.in}

\author{Tousik Samui}
\email{tousiksamui@gmail.com}
\affiliation{Regional Centre for Accelerator-based Particle Physics, 
Harish-Chandra Research Institute, HBNI, 
Chhatnag Road, Jhusi, Prayagraj (Allahabad) 211\,019, India}


\begin{abstract}
The search for new physics at high energy accelerators has
been at the crossroads with very little hint of signals
suggesting otherwise. The challenges at a hadronic machine
such as the LHC is compounded by the fact that final states are
swamped with jets which one needs to understand and unravel.
A positive step in this direction would be to separate the
jets in terms of their gluonic and quark identities, much in
a similar spirit of distinguishing heavy quark jets from
light quark jets that has helped in improving searches for
both neutral and charged Higgs bosons at the LHC. In this
work, we utilise this information using the jet substructure
techniques to comment on possible improvements in
sensitivity as well as discrimination of new resonances in
the all hadronic mode that would be crucial in pinning down
new physics signals at HL-LHC, HE-LHC and any future 100 TeV
hadron collider.
\end{abstract}

\maketitle

\section{Introduction}
Resonance search is one of the most simple and direct ways
to establish the presence of new particles at a collider.
With high centre-of-mass energies available at hadronic
colliders such as the Large Hadron Collider (LHC), it serves
as a perfect playground for resonance searches of massive
particles predicted in many beyond the standard model (BSM)
scenarios. The LHC energies and available integrated
luminosity can help search for these exotic particles beyond
4-5 TeV. The envisaged 100 TeV hadron 
machine\,\cite{Arkani-Hamed:2015vfh} will expectedly
improve this range by nearly a factor of magnitude and more
so for strongly interacting particles. However, finding a
signal for such strongly interacting heavy particles which
dominantly decay to either quarks\footnote{Here quark
represents both quark and antiquark.}, gluons or both
against the huge QCD background is very difficult. The
major challenge at a hadronic machine such as the LHC is
compounded by the fact that final states are swamped with
jets which one needs to understand and unravel in order to
extract signal for new physics in the all hadronic final
state. However, there are techniques to reduce these huge
QCD backgrounds from the signal. There already exist
searches for heavy resonances in the dijet channel by
CMS\,\cite{Sirunyan:2018xlo,Sirunyan:2016iap,
Khachatryan:2015dcf,Khachatryan:2016ecr,Khachatryan:2015sja,
Chatrchyan:2013qha} and ATLAS\,\cite{Aaboud:2018fzt,
Aaboud:2017yvp,ATLAS:2015nsi,Merten:2014wna}. The
experimental collaborations provide an upper limit at 95\%
C.L. on the cross section for heavy resonance production and
its decay to three different decay channels,
{\it viz.} $qq$, $qg$, and $gg$, separately. Although they
do not distinguish between quark or gluon jets explicitly in
the final state, they use the line-shape information
of the resonance peak to set individual limits\,\cite{Harris:2011bh} 
for the three different types of final state. In this work, we use jet
substructure techniques in studying several exotic
resonances in the dijet final state and show that in
addition to the line-shape, if we can distinguish between
the flavour\footnote{Here flavour means light
quark/antiquark and gluon jets.} of the jets, existing and
future limits on the cross section for heavy resonances can
be significantly improved. Thus a good knowledge of
discrimination between quark and gluon jets will not only
help in improving the resonance searches at the HE-LHC and 
future 100 TeV machine but also help in pinning down the new
physics scenario by giving hints on the interaction Lagrangian.

The recent developments in the study of jet substructure
provide us with one of the best ways to gather more
information from a collider event with large hadronic
activity\,\cite{Sirunyan:2017nvi,Sirunyan:2019vxa,Sirunyan:2019sgo,Sirunyan:2017dnz,Aaboud:2019zxd}. The use of jet substructure is becoming popular
day by day due to the recent developments in both
theoretical and experimental understanding of physics inside
the jets, where perturbative and non-perturbative effects
dominate in different regions in the energy scales. It has
been shown that theoretical understanding about the
substructure of a jet can even allow us to differentiate
between quark-initiated and gluon-initiated jets to a
certain extent\,\cite{Larkoski:2019nwj,Gallicchio:2012ez,
Gras:2017jty,Aad:2014gea,ATLAS-CONF-2016-034,
CMS-PAS-JME-13-002,CMS-PAS-JME-16-003,CMS-DP-2016-070,
Bright-Thonney:2018mxq,Bhattacherjee:2016bpy,
Gallicchio:2011xq,Sakaki:2018opq} due to the difference in
their radiation pattern inside the jet. Further improvements
in this direction can be possible with the use of machine
learning techniques which along with the substructure
picture of the jets makes the discrimination even more
robust and allows us to enhance the discrimination power
between quark and gluon jets\,\cite{Komiske:2016rsd,
Metodiev:2017vrx,Cheng:2017rdo,Luo:2017ncs,Fraser:2018ieu,
Kasieczka:2018lwf,ATL-PHYS-PUB-2017-017,Lee:2019ssx}. This
quark and gluon tagging can therefore be used to improve the
search for resonances in several BSM physics scenarios at
the colliders.
 
While discrimination between a quark and a gluon jet would
perhaps be a novel approach to identify certain BSM
scenarios, in this work we take a slightly different
approach. Instead of distinguishing a quark jet from a gluon
jet on a jet-by-jet basis, we take an event containing
several jets as a whole and the analysis is done on the
event variables as well as on the jet substructure
observables. The jet substructure observables chosen for
quark-gluon discrimination will have different probability
distributions for quark and gluon jets. In general, all the
jets in a set of events produced at the collider are not
entirely quark-initiated or gluon-initiated jets, but a
certain fraction of them are quark jets or gluon jets
contributing to the events. If these fractions are different
for signal events from the background events, there will be
some degree of distinguishing power between signal and
background. Through this work, we try and show that even if
we do not tag the jets in a particular event as quark or
gluon jets, we can still conclude with some confidence on
whether the event is a signal-like event or a
background-like event.

The fraction of quark or gluon jets in a set of events,
which we talked about in the last paragraph, can be a fixed
fraction generated in every single event or it can be an
overall fraction from a set of events. For example, if a
strongly interacting heavy fermion (e.g. excited quarks
$q^*$) decays to two light-flavoured jets, then one expects
that there will be 50\% quark jets and 50\% gluon jets in
each event and hence the fraction will be 50\% in the whole
set of events for both quark and gluon jets. On the other
hand, if a heavy coloured boson (e.g. a coloron) decays to
two light-flavoured jets then it will either decay to a pair of
quark jets or to a pair of gluon jets. Hence the overall fraction
of quark jets and gluon jets in the collection of events
will be determined by the relative coupling of the heavy
boson with a quark pair and a gluon pair. However, note that
such a fraction in the events could also originate from
non-resonant dijet or multijet production processes at a
collider. In fact, the Standard Model (SM) dijet background
is a non-resonant production of two light flavoured jets.
The probability distribution for jet substructure
observables will nearly be the same for a given fraction
irrespective of whether it originates from every single
event or from a collection of events. Thus closer scrutiny
of the events with jet tagging, comparing quark-initiated
and gluon-initiated jets would be crucial in understanding
the heavy resonances.\\

In this work, we highlight the improvements that
can be achieved in dijet resonant searches by the use of jet
substructure techniques. Similar studies have been done
previously, 
where in Ref.~\cite{Ozturk:2014hga} the jet substructure observables 
have been suggested to improve the searches 
while Refs.~\cite{Chivukula:2014pma,Simmons:2018qgn} use color
discriminant variable and jet energy profile as the means of
improving the sensitivity in the dijet channel. 
However, all three studies considered parton showered events only. 
Our work is more along the lines presented in Ref.~\cite{Ozturk:2014hga} but with 
substantial  improvements in the analysis.
We include the SM QCD background and use a fast detector simulation 
and account for the correction to jet energy scales in the
computation of the event variables, similar to the procedure
followed by experimental collaborations. Moreover, we use an 
additional set of observables which are considered to be
useful in CMS and ATLAS analyses~\cite{Aad:2014gea,
ATLAS-CONF-2016-034,CMS-PAS-JME-13-002,CMS-PAS-JME-16-003,
CMS-DP-2016-070}. We also show how one can use jet substructure 
technique and the boosted decision tree (BDT) multivariate method to distinguish 
different type of resonances in the dijet channel.

\section{Dijet Resonances}
In this study we consider three different kind of dijet
resonances, each of which gives us one of the three types of
dijet signals: $qq$, $qg$, or $gg$. For resonances leading
to $qq$ final state, each of the jets will be quark-like
while it will be gluon-like for $gg$ final state. On the
other hand, for $qg$ final state, its properties will be
an admixture of both quark and gluon-like. Hence, if we consider
jet substructure observables of the hardest jet (or second
hardest jet), we expect their distributions (observables sensitive to jet-types) 
to be different for these three cases. The distributions for observables of $qq$ 
will be far apart from those of $gg$ final state while for the $qg$ final state,
the distribution is expected to show an overlap with above two. 

\begin{table}[!h]
\begin{center}
\begin{tabular}{|l|c|c|r|}
\hline
Model   & Lagrangian & Parameter Value & $\sigma_{\rm total}$ [fb] \\[2pt]
\hline
Triplet Diquark & $g_s f_s \epsilon^{a b c} \phi^\dagger_a\,\bar{d}_b u^C_c$ & $f_s=0.04$ & 101.1~~ \\[2pt]
\hline
Coloron & $g_sf_s\,{G'}_\mu^a\,\bar u \gamma^\mu T_a u$ & $f_s = 0.09$ & 122.0~~ \\[2pt]
\hline
Excited Quark & $g_s \frac{f_s}{2\Lambda} \bar{U}^*[\gamma^\mu,\gamma^\nu]T_a u\,G_{\mu\nu}^a + h.c.$ & $\frac{f_s}{\Lambda}=1.5\times10^{-5}$  GeV$^{-1}$ & 219.4~~ \\[2pt]
\hline
Spin-3/2 & $i g_s \frac{f_s}{\Lambda} \bar{\psi}_\mu (g^{\mu\nu} + A \gamma^\mu\gamma^\nu) \gamma^{\sigma}T^a u G^a_{\nu\sigma} + h.c. $        & $\frac{f_s}{\Lambda}=4.1\times10^{-5}$ GeV$^{-1}$, $A=0$  & 222.3~~ \\[2pt]
\hline
RS Graviton & $\kappa\, h_{\mu\nu} T^{\mu\nu}$         & $\kappa = 5\times 10^{-5}$ GeV$^{-1}$ &  179.4~~ \\[2pt]
\hline
Color Octet Scalar & $g_s\,\frac{f_s}{\Lambda}\,d_{abc}\,\phi^a\,G^{b\,\mu\nu} G_{\mu\nu}^c$         & $\frac{f_s}{\Lambda}=5.8\times10^{-5}$ GeV$^{-1}$  & 493.9~~ \\[2pt]
\cline{1-4}
\end{tabular}
\caption{Examples of resonant particles with the production
cross section at 13~TeV LHC for each resonant particle at
2~TeV.}\vspace{-16pt}
\label{tab:cross-sections}
\end{center}
\end{table}

In Table \ref{tab:cross-sections}, we list a few important BSM
models and their relevant part of Lagrangians which would yield 
the dijet signal processes in these models,
({\it e.g.} see Refs.\,\cite{Karabacak:2012rn,Dicus:2012uh,
Das:2015lna}). Note that we have classified the choice of models based 
on the intrinsic spin of the resonance in the dijet signal. A more 
comprehensive list of all possible coloured particles that give a dijet 
resonance at the LHC can be found in Ref.\,\cite{Han:2010rf}. We also list the  
production cross sections for the resonant exotic particle 
at 13 TeV LHC for the specific values of the parameters listed 
in Table \ref{tab:cross-sections} for each model. The parameter 
values have been chosen such that the cross sections remain below the
existing upper limit from the dijet resonance search by
CMS\,\cite{Sirunyan:2018xlo}. For further analyses, we restrict ourselves 
to only three specific models (as no significant difference was observed 
in the current analysis between models where the resonance differed in spin but 
gave the same dijet final state). We therefore chose 
the coloron model\,\cite{Hill:1991at,Hill:1993hs,
Chivukula:1996yr} for $qq$ resonance, color octet scalar
model\,\cite{Hill:2002ap,Gresham:2007ri,Gerbush:2007fe}
for $gg$ resonance, and excited quark
model\,\cite{Cabibbo:1983bk,DeRujula:1983ak,Kuhn:1984rj,
Baur:1987ga,Baur:1989kv} for $qg$ resonance.

\section{Jet Substructure Observables}
For the jet substructure analysis we first list out the
criterion and observables we are interested in. We classify
our requirements based on the event variables and therefore
we choose only the simple variables, {\it viz.} $p_T$ and
$\eta$ of the leading and sub-leading jet (arranged
according to larger transverse momenta) and their angular
separation $\Delta R(j_1,j_2) =\sqrt{(\Delta\eta)^2 +
(\Delta\phi)^2}$ defined between the two jets. For the jet
substructure observables, we refer to the two types of
observables as pointed out in
Ref.\,\cite{Gallicchio:2012ez,CMS-PAS-JME-13-002,CMS-PAS-JME-16-003}, {\it viz.}
discrete and continuous observables. The most important
ones are:
\begin{itemize}
	\item Particle and charged particle multiplicity inside a jet\,\cite{Gallicchio:2012ez}.
	\item Les\,\,Houches\,\,Angularity: ${\rm LHA}\,=~\sum_{i\in J} \dfrac{p_{T_i}}{p_{T_J}} \sqrt{\dfrac{\Delta R(i,J)}{R}}$ \cite{Badger:2016bpw,Gras:2017jty}.
	\item Girth: $g =  \sum_{i\in J} \frac{p_{T_i}}{p_{T_J}}\left(\dfrac{\Delta R(i,J)}{R}\right)^2$\,\cite{Gallicchio:2012ez}. 
	\item Width: $w =  \sum_{i\in J} \frac{p_{T_i}}{p_{T_J}}\left(\dfrac{\Delta R(i,J)}{R}\right)$\,\cite{Catani:1992jc,Rakow:1981qn,Ellis:1986ig}.
	\item Two-point energy correlation variables:\\$e_\beta = \sum_{i>j\in J} \frac{p_{T_i}p_{T_j}}{p_{T_J}^2}\left(\dfrac{\Delta R(i,j)}{R}\right)^\beta$\,\cite{Larkoski:2013eya}.
	\item $p_T^2$ weighted jet minor axis ($\sigma_2$) with respect to the jet axis in $\eta-\phi$ plane\,\cite{Gallicchio:2010sw,CMS-PAS-JME-13-002}.
    \item $p_T D = \dfrac{\sqrt{\sum_{i\in J}p_{T_i}^2}}{\sum_{i\in J} p_{T_i}}$\,\cite{CMS-PAS-JME-13-002}.
\end{itemize}
where $R$ is the radius parameter of the jet and, in the
subscripts, $i,j$ represent the constituents of the jet
while $J$ represents the jet.

We know from first principle as well as from data
collected in collider experiments that radiation pattern
inside a gluon jet differs from that of the light-quark
jets. Due to the difference in radiation pattern, many jet
substructure observables have been proposed in the
literature to discriminate between quark and gluon jets.
Though our primary aim in this work is not distinguishing
a quark jet from a gluon jet, our analysis procedure is
still guided by similar jet substructure observables that
help in quark-gluon discrimination. The choice of the
observables can be explained as follows. 
\begin{itemize}
\item Gluon fragments more than a quark due to its colour
      factor (colour factor ratio is $\frac{C_A}{C_F}=
      \frac{9}{4}$), which results in higher particle
      multiplicity in a gluon jet compared to a quark jet.
\item Charged particle multiplicity is also more in case of
      gluon jets than in quark jets.
\item Gluon fragmentation function is softer than that of a
      quark, {\it i.e.} constituents of gluon jets tend to
      be softer than a quark jet. This means $p_T D \to 0$
      in the case of a gluon jet while $p_TD \to 1$ in case
      of quark jet.
\item Gluon jets are less collimated compared to quark jets.
      This gives wider radius for gluon jet with respect to
      quark jet. So, if the shape of a jet is approximated
      to an ellipse in $\eta-\phi$ plane, a gluon jet tends
      to have longer minor axis than a quark jet.
\item Combination of the last two points also tends to give
      higher values for girth, width, LHA and two-point
      energy correlation variables ($e_\beta$) for gluon
      jets compared to quark jets.
\end{itemize}

\begin{figure*}
\begin{center}
	\includegraphics[width=0.9\textwidth]{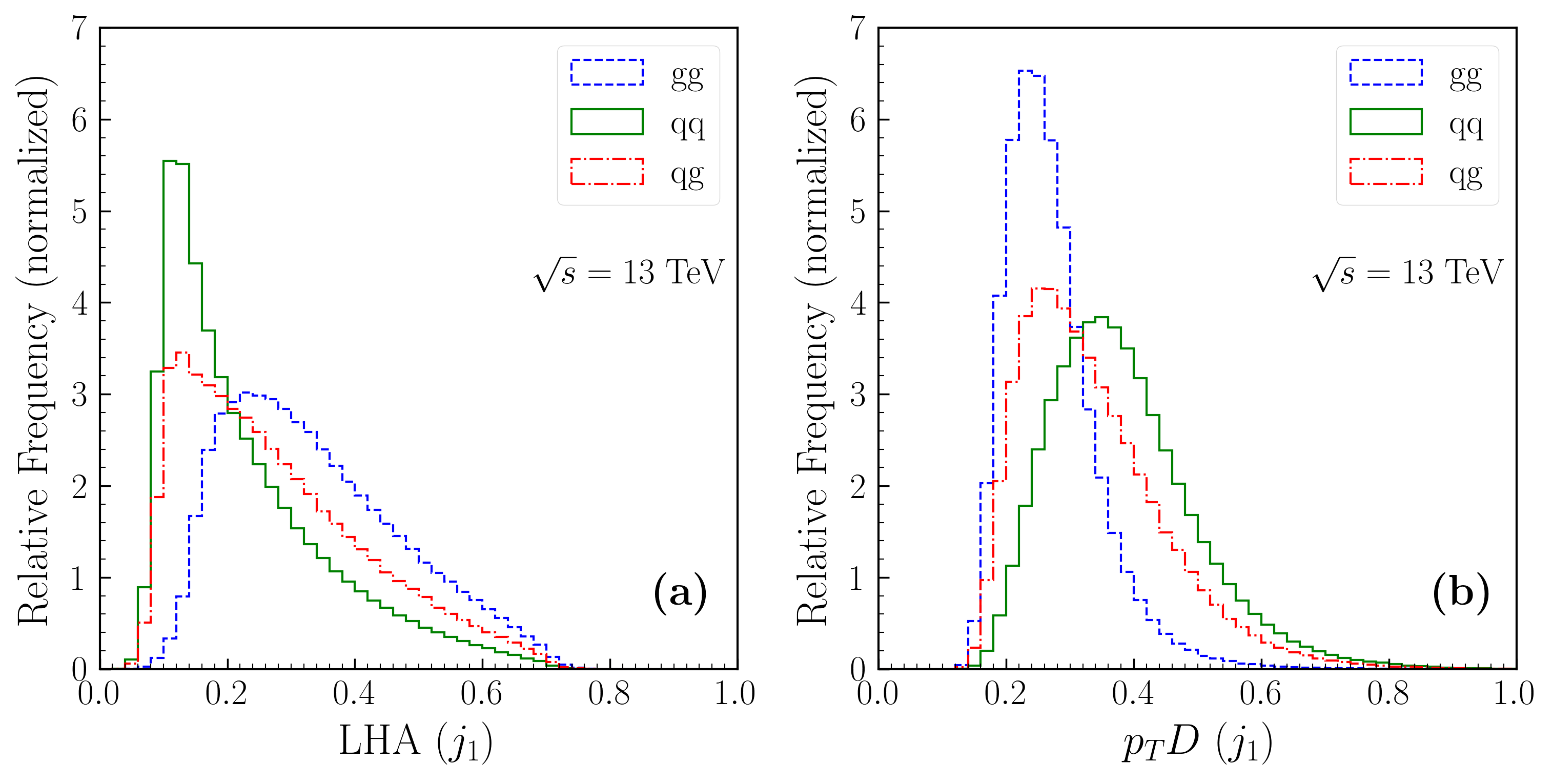}
\end{center}
\vspace{-20pt}
\caption{Distribution of jet substructure observables for SM
	$gg$ events (blue dashed), $qq$ events (green solid), and
	$qg$ events (red dot-dashed). The distributions shown are
	for (a) Les Houches Angularity (LHA) and for (b) $p_T D$.}
\label{fig:jss-obs}
\end{figure*}

To illustrate the difference in the distribution of jet
substructure observables between quark and gluon jets, in
Fig.~\ref{fig:jss-obs} we show area normalized distributions
for the two jet substructure observables, both for quark
(green solid) and gluon (blue dashed) jets in $(qq)$ and
$(gg)$ subprocesses in SM respectively. The jet
substructure observables are calculated from the leading
jets in the subprocesses. We can clearly see that the
distribution for a quark jet is quite different from that of
a gluon jet. However, for a mixed sample of quark and gluon
jets, the distribution will be smeared out. The same two
observables are plotted (red dot-dashed) in
Fig.~\ref{fig:jss-obs} for the leading jet in $qg$ events.
A selection criteria of $p_{T_J} > 500$~GeV and $|\eta|<2.5$
for both the jets was imposed on the event sample. The
anti-kt algorithm with radius parameter $R=0.4$ was used
for jet clustering. The distribution of the jet substructure
observables shown in Fig.~\ref{fig:jss-obs} are for Les
Houches Angularity (LHA) and for $p_TD$. This difference
in distribution among quark, gluon and admixture of quark
and gluon will form the basis of the analyses to follow.

\section{Analysis and Results}
To facilitate this study, Universal FeynRules Object
(UFO) \cite{Degrande:2011ua} files were generated
corresponding to the Lagrangian listed in
Table~\ref{tab:cross-sections} using {\tt
FeynRules2.0}\,\cite{Christensen:2008py, Alloul:2013bka}.
The mass of the heavy resonances have been taken to be
2~TeV for all the three types of resonances. Parton-level
dijet final states from the decay of resonant particles
were then simulated at $\sqrt s = 13$~TeV using the UFO
files with the help of {\tt
MadGraph5} \cite{Alwall:2014hca} with a hard $p_T$ cut to
populate the relevant phase space with enough events. The
events generated with  $p_T > 700$ GeV for both the jets
(parton-level) from MadGraph5 were parton showered and
hadronized using {\tt Pythia8235}\,\cite{Sjostrand:2014zea}
with the default tune implemented in Pythia. These showered
events were then passed on to {\tt
Delphes3.4.2}\,\cite{deFavereau:2013fsa} for detector
simulation. An average of 25 pile-ups have been merged with
the generated events inside the Delphes. Jets were clustered
using the anti-kt algorithm with radius $R=0.4$ from the
Particle Flow outputs from Delphes with the help of
{\tt FastJet 3.3.2} \cite{Cacciari:2011ma}. In order to
reduce the effects of contamination from Underlying Event
(UE) and pile-up (PU), many different taggers and
groomers have been proposed in the literature {\it e.g.}
trimming\,\cite{Krohn:2009th}, pruning\,\cite{Ellis:2009su,
Ellis:2009me}, mass drop tagger\,\cite{Butterworth:2008iy}
and soft drop\,\cite{Larkoski:2014wba} groomer. In this
work, we use the soft drop groomer which is a IRC safe
groomer. After jet clustering, the soft drop groomer was
used to groom away the UE and the PU with soft drop
parameter $\beta=1.0$ and $z_{\rm cut}=0.1$. Both the
groomed jets from each event are required to have $p_T$
more than 700 GeV.

Next, we perform {\it jet energy correction} on the groomed
jets, {\it i.e.} correct the detector-level jet to the
particle-level jet following the procedure prescribed in
Ref.~\cite{Aad:2011he}. We correct the jet four-momenta of
the groomed jets by calibrating groomed jets with respect to
the Monte Carlo (MC) truth jets. For this purpose, SM dijet
samples were generated using {\tt
MadGraph5}\,\cite{Alwall:2014hca} in $p_T$ and $\eta$ bins.
The MC truth jets were then constructed from the stable
particles after showering the dijet events using {\tt
Pythia8235}\,\cite{Sjostrand:2014zea}. The groomed jets were
constructed from the Particle Flow outputs from Delphes
following the same procedure described in the previous
paragraph. The algorithm and parameters for jet clustering
and soft drop groomer were taken to be the same in this case
also. The groomed jets have been constructed after merging
an average of 25 pile-up event, while no pile-up events have
been considered in the case of MC truth jet sample. For the
calibration, we imposed a matching condition of $\Delta R <
0.25$ between MC truth jets and groomed jets. The events
containing at least two matching jets were further
considered for the calibration. For each event, we then
calculate response as
\begin{equation}
R\left(p_T^{\rm groomed}\right) = \frac{p_T^{\rm groomed}}{p_T^{\rm truth}}
\end{equation}
for each of the two jets separately. In each $p_T$ and
$\eta$ bin, we next compute average of $p_T^{\rm groomed}$
$\left(\overline{p_T^{\rm groomed}}\right)$ and mean
response $\Big(\overline R \Big)$ as a function of
$\overline{p_T^{\rm groomed}}$. In each $\eta_j$ bin, the
response function $\mathcal{R}_{\eta_j}\!\left(p_T^{\rm
groomed}\right)$ is obtained from the fit of
$\left(\overline{R_i}, \overline {p_{T_i}^{\rm
groomed}} \right)$ data, where $i$ runs over $p_T$ bins. The
fitting function $\mathcal{R}_{\eta_j}$ is parametrized as
\begin{equation}
\mathcal{R}_{\eta_j}\!\left(p_T^{\rm
groomed}\right) = a_0 + a_1 \ln\left(\overline {p_T^{\rm groomed}}\right),
\end{equation}
where $a_0$ and $a_1$ are fitting parameters. The response
function has been computed for leading and sub-leading
groomed jets separately. The corrected four momenta ($p$)
of the groomed jets are then obtained by multiplying with
the inverse of the response function
\begin{equation}
p^{\rm corrected}_{\eta_j} = \frac{p^{\rm groomed}_{\eta_j}}{\mathcal{R}_{\eta_j}\!\left(p_T^{\rm
groomed}\right)} \,\, .
\end{equation}

\begin{figure}
\begin{center}
	\includegraphics[width=0.5\textwidth]{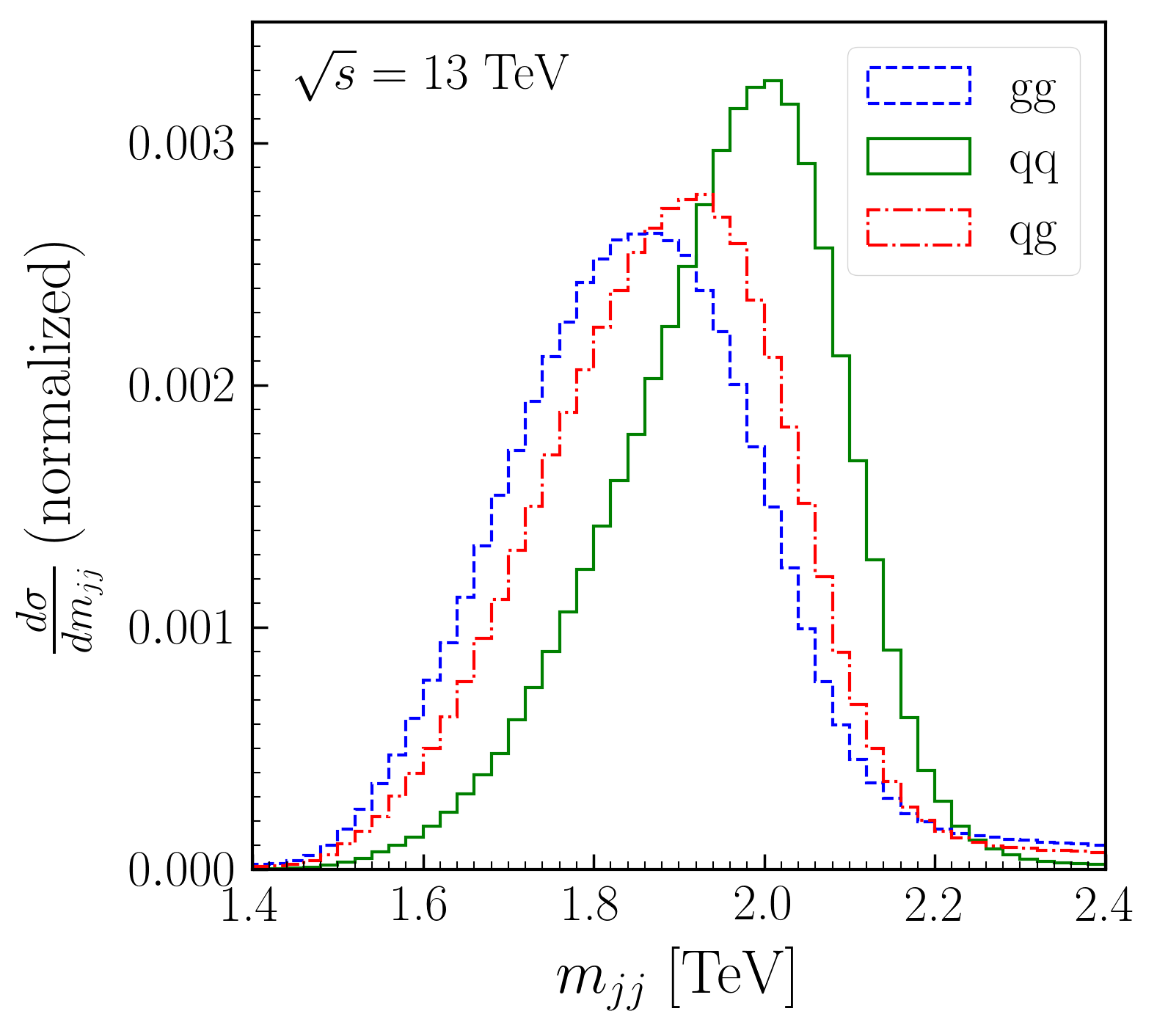}
\end{center}
\vspace{-24pt}
\caption{Distribution of dijet invariant mass for three
different types of signals, {\it viz.} $gg$, $qg$ and $qq$
resonance signals.}
\vspace{-8pt}
\label{fig:mjj}
\end{figure}

With the four momenta corrected groomed jets, we then
reconstruct the mass of the heavy resonance by taking the
invariant mass of the dijet system. The normalized
distribution of dijet invariant mass is plotted in
Fig.~\ref{fig:mjj}. The dijet invariant mass distribution
is one way of looking at the parton content of the jets
making up the resonance. It is quite clear that the
line-shapes of the resonant mass are different for
$qq,\ qg,$ and $gg$ resonances as also mentioned in
Ref.\,\cite{Sirunyan:2018xlo}. The CMS collaboration has
actually used the line shape information to put 95\% C.L.
upper bound on the three above mentioned configurations of
the dijet resonances\,\cite{Sirunyan:2018xlo}. A quick
comparison of the line-shape plot with that of
Ref.\,\cite{Sirunyan:2018xlo} finds that the $m_{jj}$
distributions are in agreement reasonably well. An upper cut
on $|\eta_j| < 2.15$ for both the jets has been imposed for
further analysis. Note that all jet substructure observables
have been constructed after the soft drop grooming
procedure. However, the four momenta, $p_T$ and invariant
mass of the dijet system are taken only after jet energy
correction.

With the generated events and variables, we then try to find
distinguishing score of the three different types of
resonances from the SM background. Since, at a hadron
collider, one cannot really avoid SM QCD background, we
consider the non-resonant SM dijet as our background in this
analysis. Note that for the SM QCD background, we shall use
the same tools and parameters discussed earlier. The dijet
background has been simulated at the leading order (LO) of
QCD. The properties of the leading and sub-leading jets of
the QCD background will again be a mixture of the properties
of pure quark and pure gluon jets. So we expect better
discrimination for $qq$ and $gg$ resonances from the SM QCD
dijet background. However, we still expect some degree of
discrimination for $qg$ resonance since the fraction of
quark and gluon will be different for the SM background from
the $qg$ resonance, where it is expected to be an admixture
of 50\% quark and 50\% gluon.

\begin{figure*}
\begin{center}
    \includegraphics[width=1.0\textwidth]{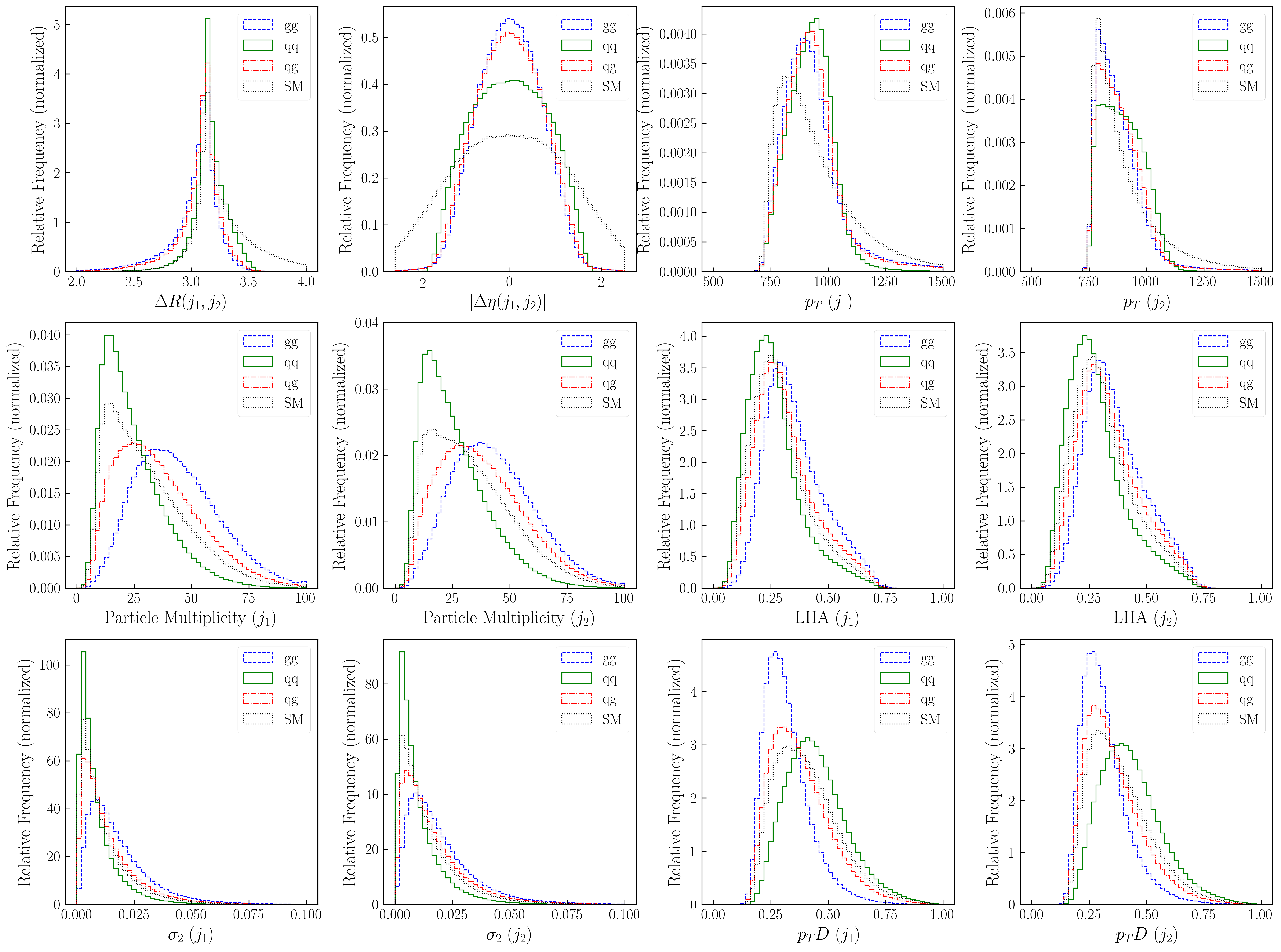}
\end{center}
\vspace{-16pt}
\caption{Normalized distribution of jet substructure
observables for the three types of signals as well as for
the SM dijet background. In all the panels, blue dashed
lines are for $gg$ resonance, green solid lines are for
$qq$ resonance, red dot-dashed lines are for $qg$ resonance,
and black dotted lines are for SM background.}
\label{fig:JSSobs}
\end{figure*}

Before going into the details of our results, it is useful
to highlight the distributions of various event variables
and jet substructure (JSS) observables. We therefore plot
the distributions of some of the important event variables
as well as JSS observables in Fig.~\ref{fig:JSSobs}, which
shows the normalized distributions of these observables for
leading ($j_1$) and sub-leading ($j_2$) jets. In all the
panels of Fig.~\ref{fig:JSSobs}, the blue dashed lines
represent the distributions of $gg$ resonance, while the
green solid lines are for $qq$ resonance and the red
dot-dashed lines are for $qg$ resonance. The normalized
distributions for the SM dijet background are also shown in
black dashed lines for all the variables. We note that the
distributions of JSS observables for $j_1$ and $j_2$
(arranged in descending order of $p_T$) are similar for
different types of signals as well as for the background.
This is expected because of the following reasons. In the
case of $gg$ ($qq$) resonance signal, both the jets are
gluon-initiated (quark-initiated) jets. In the case of $qg$
resonance signal, the percentage of quark-initiated and
gluon-initiated jets in $j_1$ is the same as in $j_2$ in a set
of events. The same argument applies for the SM dijet
background as well.

One may now proceed on to perform a cut based analysis to
maximize the different types of signals with respect to the
background using the shapes of the distributions. However, a
multivariate analysis (MVA) further enhances the signal with
respect to background by providing a classifier variable
after combining all the input variables in an optimized way.
In this study, we used BDT
classifier variable for the rest of the analyses to follow.
For the implementation of BDT, the {\tt
TMVA2.0}\,\cite{Hoecker:3039H} package which is built-in in
{\tt Root6}\,\cite{Brun:1997pa,root:v6-16} was used. At this
point, we note that though the resonance masses of all the
three different types of resonances have been taken to be
the same (2~TeV), the peak positions and spreads of the
invariant mass distribution, as shown in Fig.~\ref{fig:mjj},
for different types of resonances are slightly different.
Hence, in order to maximize the effect of BDT analysis, we
put the following cut on the invariant mass of the different
types of resonances\footnote{Note that we want the BDT to
exploit more information from the JSS variables and focus on
the backgrounds that are lying under the signal peak and can
not be easily suppressed by looking at dijet mass only.
Though we propose a full search analysis in an experiment to
construct mass dependent BDT discriminators, here we present
only a case study considering only a single mass point at 2
TeV. This search strategy however can be replicated for each
considered mass point in a real experimental analysis.}
before passing them on to the BDT analysis.
\begin{itemize}
\item $1580 < M_{gg} < 2100$
\item $1680 < M_{qq} < 2180$
\item $1600 < M_{qg} < 2120$
\end{itemize}

\begin{table}[!h]
\begin{center}
\begin{tabular}{?l?l?c|c?c|c?c|c?}
\cline{1-8} \multicolumn{2}{?c?}{} &  \multicolumn{6}{c?}{} \\[-16pt]
\multicolumn{2}{?c?}{\multirow{3}{*}{\bf ~Variables}}  & \multicolumn{6}{c?}{Variable Importance (in \%)} \\[2pt]
\cline{3-8} \multicolumn{2}{?c?}{} & \multicolumn{2}{c?}{} & \multicolumn{2}{c?}{} & \multicolumn{2}{c?}{} \\[-16pt]
\multicolumn{2}{?c?}{~} &  \multicolumn{2}{c?}{$gg$} & \multicolumn{2}{c?}{$qq$} & \multicolumn{2}{c?}{$qg$}\\[2pt]
\cline{3-8} \multicolumn{2}{?c?}{} & & & & & & \\[-16pt]
  \multicolumn{2}{?c?}{~} &  Jet 1 & Jet 2 & Jet 1 & Jet 2 & Jet 1 & Jet 2\\[2pt]
\cline{1-8} & & \multicolumn{2}{c?}{} & \multicolumn{2}{c?}{} & \multicolumn{2}{c?}{} \\[-16pt]
 \multirow{5}{*}{\quad\rotatebox[]{90}{\bf Event\qquad\quad}~~\rotatebox[]{90}{\bf Variables\quad\qquad}\quad} & ~$m_{jj}$               & \multicolumn{2}{c?}{7.38} & \multicolumn{2}{c?}{9.41} & \multicolumn{2}{c?}{8.81} \\[2pt]
\cline{2-8} & & \multicolumn{2}{c?}{} & \multicolumn{2}{c?}{} & \multicolumn{2}{c?}{} \\[-16pt]
 & ~$\Delta R (j_1,j_2)$     & \multicolumn{2}{c?}{9.16} & \multicolumn{2}{c?}{7.03} & \multicolumn{2}{c?}{8.41} \\[2pt]
\cline{2-8} & & \multicolumn{2}{c?}{} & \multicolumn{2}{c?}{} & \multicolumn{2}{c?}{} \\[-16pt]
 & ~$|\Delta \eta(j_1,j_2)|$        & \multicolumn{2}{c?}{5.34} & \multicolumn{2}{c?}{5.67} & \multicolumn{2}{c?}{5.67} \\[2pt]
\cline{2-8} & & & & & & & \\[-16pt]
 & ~${p_T}$                  & 5.70 & 4.87 & 6.24 & 8.50 & 7.68 & 6.16 \\[2pt]
\cline{2-8} & & & & & & & \\[-16pt]
 & ~Energy                   & 5.12 & 5.09 & 4.19 & 4.37 & 4.56 & 4.56 \\[2pt]
\cline{2-8} & & & & & & & \\[-16pt]
 & ~$\eta$                   & 4.47 & 4.03 & 4.05 & 4.29 & 4.50 & 4.48 \\[2pt]
\cline{1-8} & & & & & & & \\[-16pt]
 \multirow{4}{*}{\quad\rotatebox[]{90}{\bf JSS\quad}~~\rotatebox[]{90}{\bf Observables\quad}\quad} & ~Particle Multiplicity~ & \quad \,11.98 \quad\quad &\quad \,9.23 \quad\quad &\quad \,8.43 \quad\quad & \quad \,7.88 \quad\quad &\quad \,9.14 \quad\quad &\quad \,8.45 \quad\quad \\[2pt]
\cline{2-8} & & & & & & & \\[-16pt]
 & ~$p_T D$                & 6.20 & 4.87 & 6.23 & 6.07 & 5.57 & 5.55 \\[2pt]
\cline{2-8} & & & & & & & \\[-16pt]
 & ~LHA                    & 5.33 & 4.86 & 5.85 & 5.75 & 5.32 & 5.16 \\[2pt]
\cline{2-8} & & & & & & & \\[-16pt]
 & ~$\sigma_2$             & 3.11 & 3.26 & 2.86 & 3.17 & 2.91 & 3.06 \\[2pt]
\cline{1-8}
\end{tabular}
\caption{List of values of {\it variable importance} defined
as the weighted fraction of the times a particular variable
has been used in the boosted decision tree training. All the
values are calculated in the training with JSS observables.}
\vspace{-16pt}
\label{tab:var-rank}
\end{center}
\end{table}

As mentioned in Refs.\,\cite{Gallicchio:2012ez,Gras:2017jty,
CMS-PAS-JME-13-002,CMS-PAS-JME-16-003}, some jet substructure observables
are better suited for quark and gluon jet tagging than
others. In our work we too perform a thorough investigation
of the jet substructure observables to find out which of the
observables are best suited for better distinguishing power.
We also note that the same jet substructure observables will
be equally important in distinguishing the different types
of final state events. To highlight this, we list the {\it
variable importance} in BDT\,\cite{Hoecker:3039H} in
Table~\ref{tab:var-rank}. The {\it variable importance} is
defined as the weighted fraction of how many times a
particular variable is used to split the decision tree
nodes. The weight is the product of the separation
gain-squared it has achieved in each split occurrence and
the number of events in the node. As we see that the most
important variables are the event variables, {\it viz.}
$m_{jj}$, $\Delta R(j_1,j_2)$ and $\Delta \eta$. However,
the importance of JSS observables are not very small.
Combination of JSS observables give rise to substantial
contribution.

\begin{figure*}[h]
\begin{center}
\includegraphics[width=\textwidth]{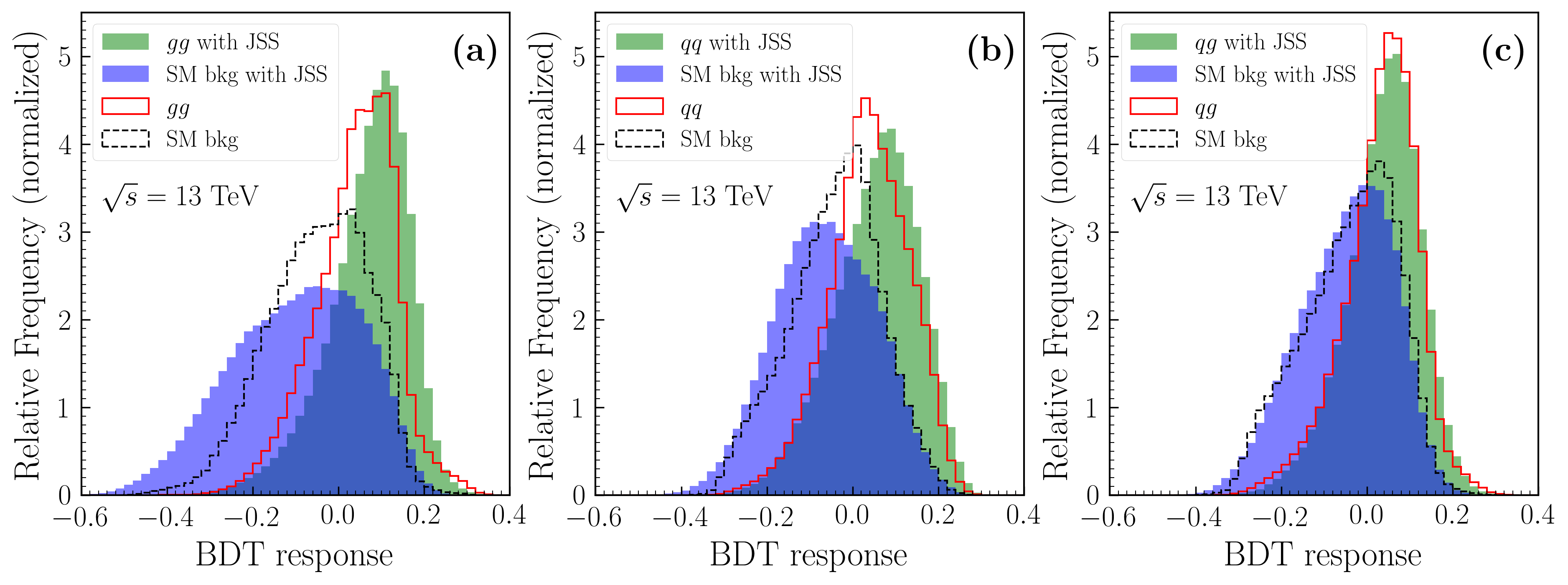}
\end{center}
\vspace{-20pt}
\caption{BDT response for the signal and the background for
three resonance cases. In all the panels, red solid (black
dashed) line represents BDT response for the signal
(background) without using the JSS observables while the
green (blue) shade represents the same with the JSS
observables. The responses shown are for (a) $gg$ resonance,
(b) $qq$ resonance, and (c) $qg$ resonance.}
\label{fig:mvavalues}
\vspace{-8pt}
\end{figure*}

In this study, our primary aim is to investigate how much
improvement one can get over the current resonant dijet
search by the use of jet substructure observables. Hence, we
optimize BDT response in two stages. In order to see that
the use of jet substructure observables help in
discriminating signal from the background, we first optimize
BDT response for the signal using only event variables,
{\it viz.} $p_T$, $\eta$ of both the jets, $\Delta
R(j_1,j_2)$, $|\Delta\eta|$, and $m_{jj}$ for the
discrimination of signal from the SM dijet background. We
then supplement the analysis by adding jet substructure
(JSS) observables, {\it viz.} particle multiplicity, LHA,
$p_T D$, $\sigma_2$ of leading as well as sub-leading jets
in addition to the simple event variables. To illustrate the
effect of JSS observables, we plot the distributions for BDT
response for different types of signals as well as for the
background in Fig.~\ref{fig:mvavalues}. In all the panels of
Fig.~\ref{fig:mvavalues}, the red solid (black dashed) line
represents the BDT response for the signal (background)
without using the JSS observables while the green (blue)
shade represents the same with the JSS observables. The
responses shown are for (a) $gg$ resonance, (b) $qq$
resonance, and (c) $qg$ resonance. The improvement of
signal-background separation can be understood from the
relative shift in the BDT response distribution for the
signal and the background in Fig.~\ref{fig:mvavalues}. As we
can see from the figure that the distribution of BDT
response for the signal (red solid) and that for the
background (black dashed) are close to each other when JSS
observables were not considered. However, after the
incorporation of the JSS observables, the two distributions
(green and blue shaded regions) are better separated. This
tells us that the use of JSS observables will further
improve the probe of the signals over the background
although the degree of improvement will be different for
different types of signals. A quick comparison among the BDT
responses tells us that the improvement will be better for
$gg$ and $qq$ resonance signals than the $qg$ signal.

\begin{figure*}[h]
\begin{center}
\includegraphics[width=\textwidth]{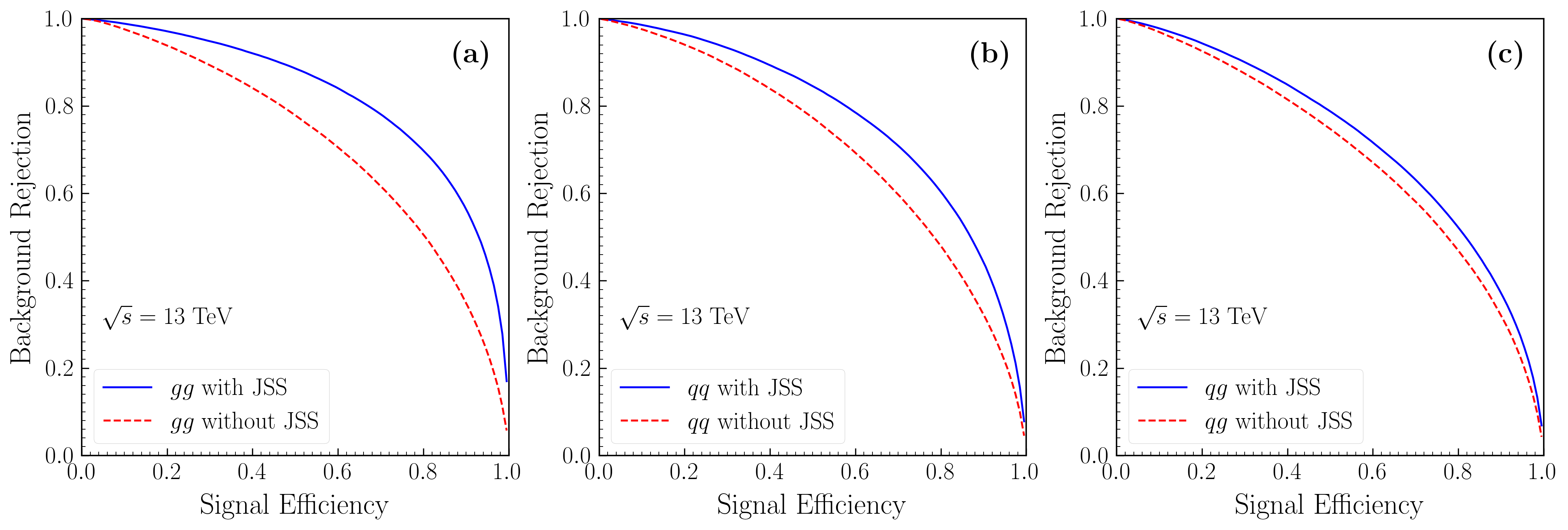}
\end{center}
\vspace{-20pt}
\caption{Illustrating ROC curves for the (a) $gg$ resonance,
(b) $qq$ resonance and (c) $qg$ resonance signal against the
SM dijet background for analyses with (blue solid) and
without (red dashed) JSS observables.}
\label{fig:ROC}
\end{figure*}

Although Fig.~\ref{fig:mvavalues} shows some degree of
improvement when JSS observables are used along with event
variables, it does not provide any quantitative assessment
of the improvement. One of the ways to compare and see the
improvement is via receiver operating characteristic (ROC)
curve. A ROC curve gives the background rejection
efficiency, {\it i.e.} the fraction of background rejected
out of the total background for a given signal efficiency
(the fraction of signal accepted out of the total signal).
For a single variable, ROC curve can easily be obtained by
sliding upper or lower cut on that particular variable and
plotting the values of background rejection efficiency
versus signal efficiency. However, for more than one
variable, the ROC curve is not unique. Since the MVA
already provides single optimized classifier variable from
more than one input variable, we can use this classifier
variable to draw ROC curves for the three different types
signals with respect to the background. We plot the ROC
curves Fig.~\ref{fig:ROC} with-JSS (blue solid) and
without-JSS (red dashed) observables for all the three
types of resonances. We can see, quite clearly, a
significant improvement if we add JSS observables to the
analysis.

\begin{figure*}[h]
\begin{center}
\includegraphics[width=0.90\textwidth]{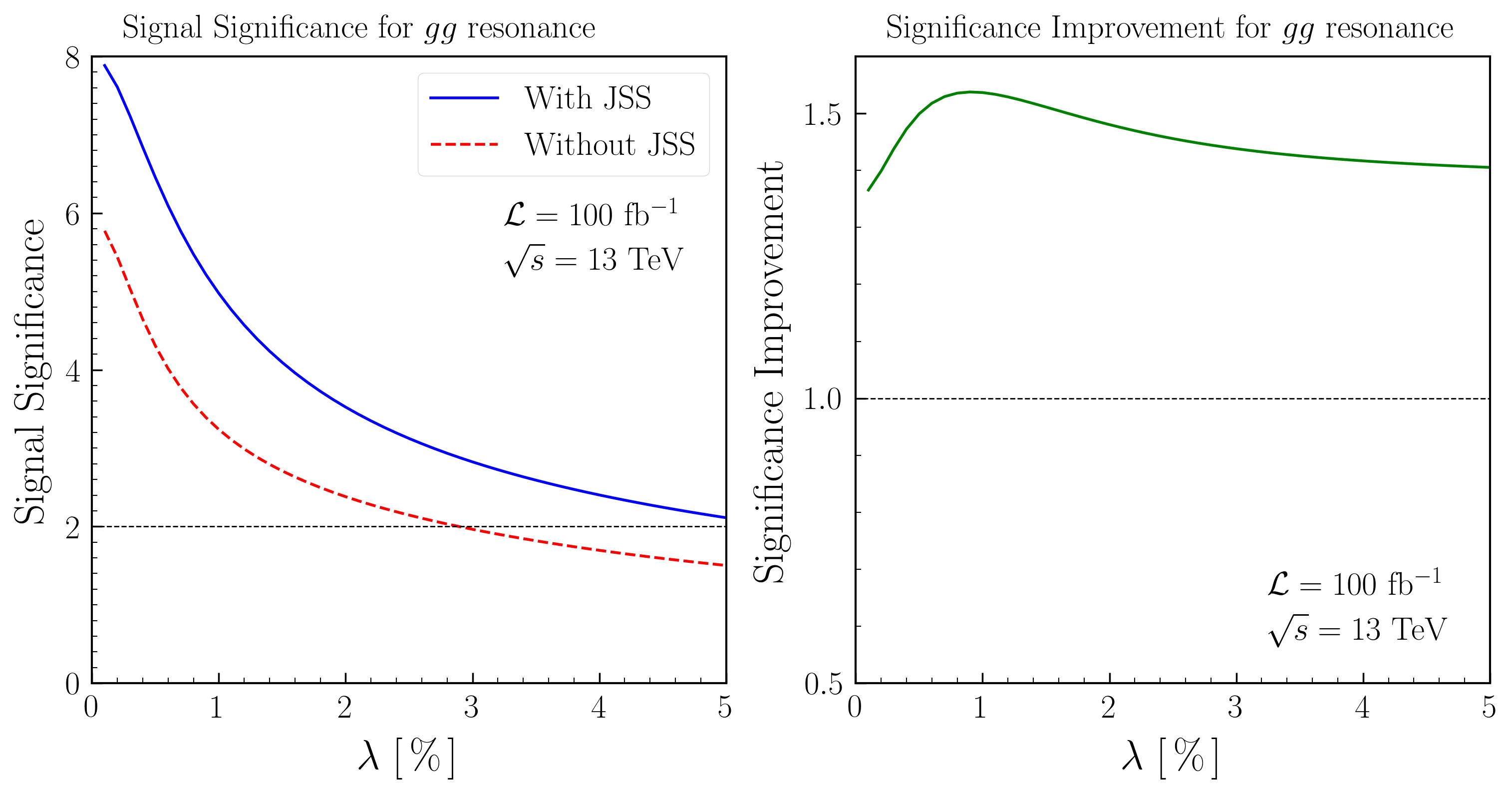}
\end{center}
\vspace{-20pt}
\caption{(left) Variation of signal significance as a
function of relative systematic uncertainty ($\lambda$) for
with-JSS (blue-solid) and without-JSS (red-dashed) case.
(right) Variation of significance improvement as a function
of $\lambda$. The variations plotted are for $gg$ resonance
signal.}
\label{fig:octet-improvement}
\vspace{-8pt}
\end{figure*}

Once the numbers of signal and background events are
estimated, we are ready to calculate signal significance by
the traditional cut and count method. In this method, one
generally counts the number of signal ($S$) and background
($B$) events after imposing a cut at a specific point in the
variable space. The cut is chosen in such a way so that the
significance, defined as $S/\sqrt{B}$, is maximized. Since
BDT response is already an optimized variable, this single
variable can be used for cut and count analysis method.
However, estimation of signal significance using the shape
of the signal and background distribution of a particular
variable gives significant improvement over the simple cut
and count method. In this work, we performed profile
likelihood ratio method to calculate signal significance. We
used profile likelihood calculator, which is implemented in
{\tt Roostats}\,\cite{Moneta:2010pm} package, after building
probability distribution functions (PDF) for the signal and
for the background using {\tt Number Counting Pdf} {\tt
	Factory}\,\cite{Cranmer:2005hi,Cranmer:1099969,
	Cousins:2008zz} from the BDT response histograms. In each
bin of the histograms, the factory incorporates relative
systematic uncertainty ($\lambda_i$) to the fraction of
events in the sideband region (defined as $\tau_i$ in
Refs\,\cite{Cranmer:2005hi,Cranmer:1099969,Cousins:2008zz}),
as follows
\begin{eqnarray}
	\tau_i = \frac{1+\sqrt{1+4\lambda_i^2}}{2\lambda_i^2 b_i}
\end{eqnarray}
where $b_i$ is the number of background and $i$ represents
the index of the bin. In this analysis, the relative
uncertainty in the background is taken to be flat, {\it
	i.e.} $\lambda_i=\lambda$ for all the bins. The variation of
the signal significance as a function of relative systematic
uncertainty ($\lambda$) is shown in the left panel of
Fig.~\ref{fig:octet-improvement} for $gg$ resonance signal.
The red-dashed line is for the analysis ``without-JSS" case
while the blue-solid line is for ``with-JSS" case. The
luminosity is taken to be 100~fb$^{-1}$ in the signal
significance calculation. From the figure, we see that the
signal significance can be improved if we use JSS
observables in addition to the event variable. To quantify
the improvement of using JSS observables, we define a
significance improvement variable as
\begin{equation}
	{\rm significance\ improvement} = \frac{\sigma^\text{JSS}}{\sigma}
	\label{eqn:SI}
\end{equation}
where $\sigma$ and $\sigma^{\rm JSS}$ represent
significances for without-JSS and with-JSS case
respectively. significance improvement is plotted as a
function of $\lambda$ in the right panel of
Fig.~\ref{fig:octet-improvement}.
We can easily say that the signal significance can be improved by about 40\% with the use of JSS observables even with systematic uncertainty higher than 2\%.

The plots in Fig.~\ref{fig:octet-improvement} consider the same values of $\lambda$ for both the
with-JSS and without-JSS case and is more of a phenomenological choice. In an experimental analysis,
with-JSS case may have additional source of uncertainty and therefore the plots for the
individual cases should be seen with respect to the absolute systematics and not relative to
each other. However, from Fig.~\ref{fig:octet-improvement}, we find that with-JSS case still performs better,
even with a $\sim$2\% additional uncertainty. Hence, one would still benefit from the study with jet substructure 
once more accurate estimates of systematics is performed in the experimental analysis.

We note that we did not carry out any explicit analysis on
the systematics. One of the major source of systematic
uncertainty could be from the determination of jet energy
scale. Though we have corrected the jet four-momenta for the
jet energy scale, we did not really estimate the uncertainty
in that. Another major source of uncertainty can be from PDF
and parton shower scale uncertainty. Also, it is not
guaranteed that the uncertainty would remain uniform in the
whole range of signal efficiency. However, our analysis is
still useful in inferring that JSS observables help improve
the performance of existing dijet analysis even in the case
of non-zero systematics.

Another important point to discuss is the calibration and estimation of systematic uncertainties in BDT data. The distributions of the input variables to the BDT can be validated in data in a background enriched region. The background enriched region, for example, can be considered to be the region in dijet mass distribution which are already excluded by experiments at 95\% CL. Similarly, the distributions of BDT output can also be validated in data in such a background enriched region. A correction can be derived and/or an additional uncertainty can be assigned if any systematic shift is observed. Similarly, the systematic uncertainties affecting the input variables can be propagated to the BDT ouput by evaluating the BDT response for up and down variations of each of the input variables.

\begin{figure*}[h]
\begin{center}
\includegraphics[width=0.9\textwidth]{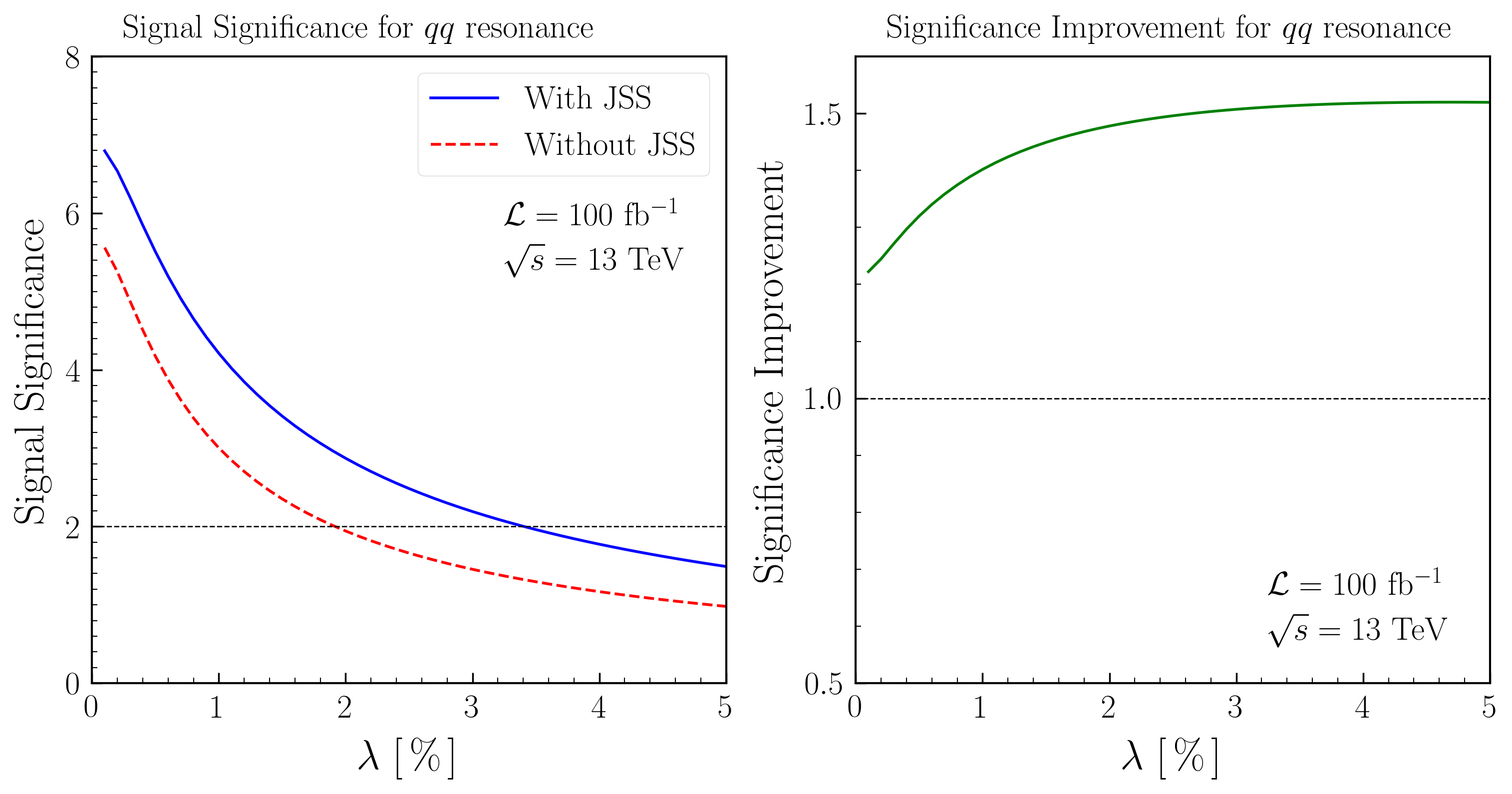}
\includegraphics[width=0.9\textwidth]{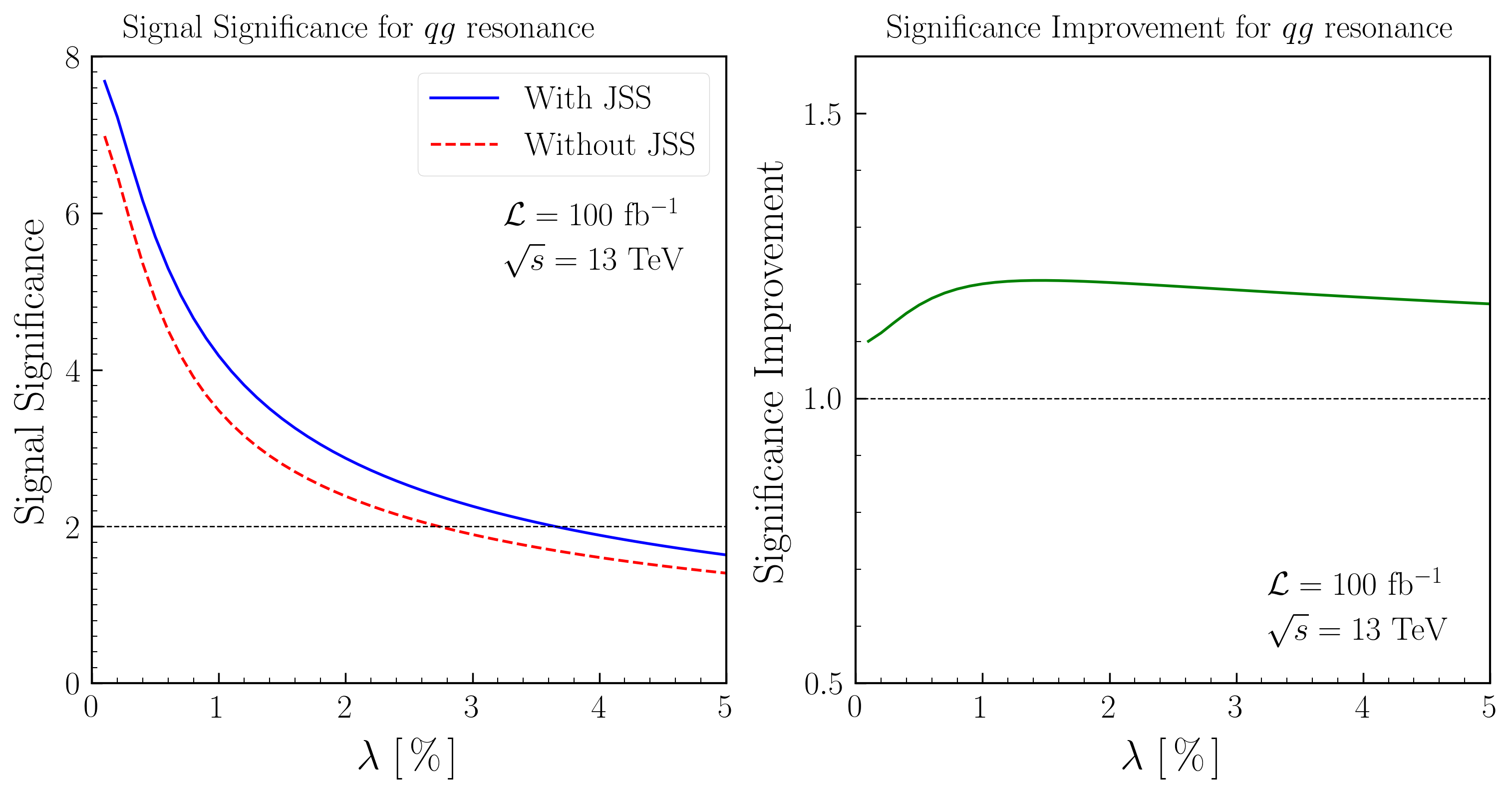}
\end{center}
\vspace{-20pt}
\caption{(left) Variation of signal significance as a
function of relative systematic uncertainty ($\lambda$) for
with-JSS (blue-solid) and without-JSS (red-dashed) case.
(right) Variation of significance improvement as a function
of $\lambda$. The variations plotted are for (top) $qq$ and
(bottom) $qg$ resonance signal.}
\label{fig:coloron-exqu-improvement}
\end{figure*}

We carried out a very similar exercise (analyses) for the
other two types of resonant configuration in the dijet, {\it
i.e.} the $qq$ and $qg$ resonances. The results are shown in
Fig.~\ref{fig:coloron-exqu-improvement}. In the left panel
of Fig.~\ref{fig:coloron-exqu-improvement}, we plotted the
variation of signal significance as a function of relative
systematic uncertainty ($\lambda$) for the $qq$ (top) and
$qg$ (bottom) resonance signals against SM dijet background.
Significance improvement as a function of $\lambda$ is
plotted in the right panel of the figure. We again observe
reasonable improvements as seen for $gg$ resonance. For $qg$
resonance the improvement is slightly depleted compared to
the other two cases. This is because both $qg$ signal and SM
dijet background has an admixture of both quark and gluon jet
properties in either jet in the dijet event. Nevertheless, we
could achieve some degree of improvement since the fraction
of quark and gluon jets are different in the signal with
respect to the background. 
A quantitative comparison of the upper limits on the product 
of cross section (partonic) and acceptance ($A$) at 95\% CL is also shown 
in Table~\ref{tab:compare} for a fixed value of the integrated luminosity of 
36 fb$^{-1}$. It is worth pointing out here that we make some approximations, such as normalizing our 
background estimates in the relevant mass bin with that of the CMS analysis and account for 
approximate acceptance values $A$,\footnote{The acceptance factors as provided in Ref.~\cite{Sirunyan:2018xlo} are taken as 57\%, 85\% and  76\% for $gg$, $qq$ and $qg$ resonances respectively. These numbers are quoted for 1.6~TeV resonance in the CMS paper.} to arrive at the upper limits shown in Table~\ref{tab:compare}.

\begin{table*}[h]
\begin{center}
\begin{tabular}{|c|c|c|c|c|}
\hline Resonance & Relative systematics & \multicolumn{2}{|c|}{Upper limit on $\sigma\times A$ (pb)} & CMS \\
\cline{3-4} type & uncertainty ($\lambda$) & ~~~With JSS~~~ & Without JSS & Limit\cite{Sirunyan:2018xlo} \\
\hline  $gg$          &  &  &  & \\[-16pt]
& $\lambda=0\%$ & 0.052 & 0.070 & 0.297 \\
\cline{2-4}           &  &  &  & \\[-16pt]
& $\lambda=2\%$ & 0.071 & 0.111 &\\
\hline    $qq$        &  &  &  & \\[-16pt]
& $\lambda=0\%$ & 0.042 & 0.052 & 0.108 \\
\cline{2-4}           &  &  &  & \\[-16pt]
& $\lambda=2\%$ & 0.058 & 0.078 & \\
\hline   $qg$         &  &  &  & \\[-16pt]
& $\lambda=0\%$ & 0.063 & 0.068 & 0.167 \\
\cline{2-4}           &  &  &  & \\[-16pt]
& $\lambda=2\%$ & 0.100 & 0.120 &\\
\hline
\end{tabular}
\end{center}\caption{Comparison of 95\% CL upper limit on $\sigma\times A$ between our analyses vs. CMS result\,\cite{Sirunyan:2018xlo} at $M=2$~TeV. The upper limits are given for two different values of relative systematic uncertainty $\lambda$. The analysis is performed for $\sqrt s = 13$~TeV and $\mathcal{L}=$ 36~fb$^{-1}$.}\label{tab:compare}
\end{table*}

\begin{figure*}
\begin{center}
\includegraphics[width=\textwidth]{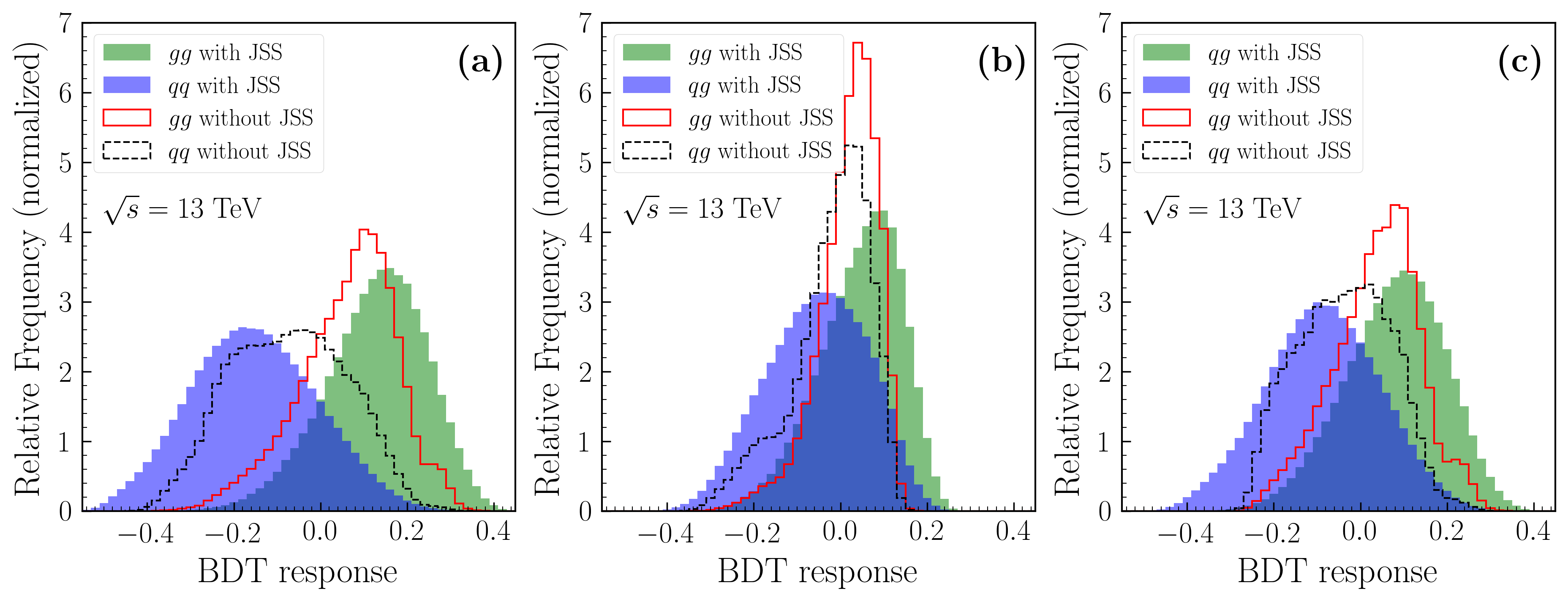}
\end{center}
\vspace{-20pt}
\caption{Normalized distribution of BDT responses for (a)
$gg$ vs. $qq$, (b) $gg$ vs. $qg$, and (c)  $qg$ vs. $qq$
signal separation. In all the panels, red solid and black
dashed lines represent the classifier responses for the
case when JSS observables are not used while green and blue
shaded regions represent the case when JSS observables are
used along with event variables.}
\label{fig:bdtresponse}
\end{figure*}

\section{Discrimination between $gg$ vs. $qg$ vs. $qq$ Resonances}\vspace{-8pt}
Once the signal is identified over the SM background, we can
expect to discriminate between different types of
resonances, {\it e.g.} discriminating $gg$ resonance from
$qq$ and $qg$ resonances. This is in similar spirit of
determining the CP properties\,\cite{Aad:2020ivc,
Sirunyan:2020sum} of the observed 125-GeV scalar at the LHC.
For the purpose of discriminating different types of
signals, the BDT is trained for one signal against the other
signal. If we consider three different types of signals, we
have three different combinations, {\it viz.} (a) $gg$ vs.
$qq$, (b) $gg$ vs. $qg$ and (c) $qg$ vs. $qq$. The BDT
responses for these three combinations are shown in
Fig.~\ref{fig:bdtresponse}. Panel (a) shows BDT responses
for $gg$ vs. $qq$ signal separation while panels (b) and (c)
show BDT responses for $gg$ vs. $qg$ signal discrimination
and $qg$ vs. $qq$ discrimination respectively. In all the
panels, red solid and black dashed lines represents the
classifier response for the case when JSS observables are
not used while green and blue shaded regions represent the
case when JSS observables are used along with event
observables. In this study we used the same set of
event variables and JSS observables as done in the analysis
of signal separation from the SM dijet background. All the
panels in Fig.~\ref{fig:bdtresponse} show similar
features, as seen in Fig.~\ref{fig:mvavalues}, {\it i.e.}
the distributions of BDT response of two resonances for
``without-JSS" case have higher overlap compared to the case
when JSS observables were used. This tells us that
different types of resonances can be better separated if we
use JSS observables in addition to the event variables.

To get a quantitative idea, we calculated separation
power\,\cite{Hoecker:3039H} for the three different
combinations, {\it viz.}~(a) $gg$ vs.~$qq$, (b) $gg$ vs.~$qg$ and (c) $qg$ vs.~$qq$. The separation power is
defined~as
\begin{equation}
\langle S^2\rangle = \frac{1}{2}\int dx \frac{\left(p(x)-q(x)\right)^2}{p(x)+q(x)}
\end{equation}
where $p(x)$ and $q(x)$ are normalized distribution functions
for the two resonances in consideration for discrimination.
The values for separation power for the BDT response
in ``without-JSS" and ``with-JSS" analysis is listed in
Table~\ref{tab:sep-pow}. We can see from
Table~\ref{tab:sep-pow} as well as from
Fig.~\ref{fig:bdtresponse} that the use of JSS observables
help us to discriminate between the different types of resonances much better. 
\begin{table*}[!h]
\begin{center}
\begin{tabular}{|c|c|c|}
\hline
 & $\langle S^2 \rangle$ (without-JSS) & $\langle S^2 \rangle$ (with-JSS) \\[2pt]
\hline
 ~~$gg$ vs. $qq$~~ & 23.8\% & 58.0\% \\[2pt]
\hline
 ~~$gg$ vs. $qg$~~ &  5.7\% & 17.8\% \\[2pt]
\hline
 ~~$qg$ vs. $qq$~~ & 11.4\% & 32.2\% \\[2pt]
\hline
\end{tabular}
\caption{Separation power between different resonance
hypotheses for without-JSS and with-JSS case.}
\label{tab:sep-pow}
\end{center}
\end{table*}

\vspace{-32pt}
\section{Summary and Outlook}
To summarize, we studied dijet resonances at the LHC and
look at the application of jet substructure techniques to
such resonances. These resonances may carry different
partonic imprints in the jets which will be driven by the
spin and  color structure of the on-shell particle produced
as a resonance. We note that JSS techniques have become an
essential part of today's collider physics and are being
utilized as a very effective tool to understand physics at
high energy colliders. Our aim of using  jet substructure in
this work is to make a statement on the improvements one can
achieve in resonant search strategies at the colliders in
the dijet final state. However, such an improvement is not
restricted to only an exclusive dijet final state but can be
applied when final states are of multijet in nature.  

Currently, experimental collaborations put 95\% C.L. upper
limit on the production cross section of different types of
resonances, {\it viz.} $gg$, $qq$, and $qg$ resonances using
only line-shape information of these types of resonances at
13~TeV collider. In this work, we attempt to make use of JSS
to improve the existing search strategies of heavy colored
resonances in dijet channel. We improve upon
earlier studies by performing a more realistic event analysis 
by including fast detector simulation and jet energy correction. 
We show how the jet substructure observables help in discriminating 
a signal for any of the heavy dijet resonances, {\it viz.} $gg$, $qq$,
and $qg$, from the SM QCD dijet background. We utilize
BDT multivariate classifier for the
discrimination. We highlight our results in the form of ROC
curves where we find considerable improvement in the ROC
curves when we use jet substructure observables in addition
to the simple event variables. We also find that the
improvement in signal significance is substantial even in
the presence of non-zero systematic uncertainties. This
suggests that, with the help of jet substructure technique,
the search for different types of heavy resonances can be
improved to a great extent at the LHC. Furthermore, we also
establish and show that distinction between different types
of resonances can also be achieved to a higher degree if jet
substructure observables are used in addition to the event
variables. The same technique can also be effectively
applied to proposed future high energy machine although
analyses with only 13~TeV has been presented in this
article.

\begin{acknowledgements}
TS and SKR  acknowledge financial support from the
Department of Atomic Energy, Government of India, for the
Regional Centre for Accelerator-based Particle Physics
(RECAPP), Harish-Chandra Research Institute. TS would like
to thank Tanmoy Mondal for useful discussion. The authors
would also like to thank Mrinal Dasgupta, and Satyaki
Bhattacharya for their insightful lectures on jet
substructure and multivariate techniques respectively,
during HEP Activity Weeks 2019 organized at Harish-Chandra
Research Institute. 
\end{acknowledgements}

\providecommand{\href}[2]{#2}\begingroup\raggedright\endgroup

\end{document}